\documentclass[runningheads]{llncs}
\DeclareUnicodeCharacter{202F}{\,}
\usepackage[utf8]{inputenc}
\usepackage{amssymb}
\usepackage{amsmath}
\usepackage{esvect}
\usepackage[T1]{fontenc}
\usepackage{graphicx,verbatim}
\begin{document}
\title{U-DFA: A Unified DINOv2-Unet with Dual Fusion Attention for Multi-Dataset Medical Segmentation}

%
\author{Zulkaif Sajjad\and Furqan Shaukat\and Junaid Mir}
\authorrunning{Sajjad et al}
%
\institute{
Department of Electrical and Electronics Engineering\\ University of Engineering and Technology Taxila, Pakistan\\
\email{zulkaifsajjad123@outlook.com}
}

    
\maketitle             
\begin{abstract}
Accurate medical image segmentation plays a crucial role in overall diagnosis and is one of the most essential tasks in the diagnostic pipeline. CNN-based models, despite their extensive use, suffer from a local receptive field and fail to capture the global context. A common approach that combines CNNs with transformers attempts to bridge this gap but fails to effectively fuse the local and global features. With the recent emergence of VLMs and foundation models, they have been adapted for downstream medical imaging tasks; however, they suffer from an inherent domain gap and high computational cost. To this end, we propose U-DFA, a unified DINOv2-Unet encoder-decoder architecture that integrates a novel Local-Global Fusion Adapter (LGFA) to enhance segmentation performance. LGFA modules inject spatial features from a CNN-based Spatial Pattern Adapter (SPA) module into frozen DINOv2 blocks at multiple stages, enabling effective fusion of high-level semantic and spatial features. Our method achieves state-of-the-art performance on the Synapse and ACDC datasets with only 33\% of the trainable model parameters. These results demonstrate that U-DFA is a robust and scalable framework for medical image segmentation across multiple modalities. The source code and pre-processed data can be accessed using the link:


\end{abstract}
\section{Introduction}
Medical image segmentation is crucial for Computer-aided Diagnosis (CAD), enabling the identification of anatomical or pathological structures in various imaging modalities. Accurate segmentation is vital for reliable diagnosis, treatment planning, and prognosis \cite{cad}. However, automating this process is challenging due to low contrast between soft tissues, high variability in anatomical and pathological structures, and the lack of annotated datasets, which complicates the modeling of relationships between structures and their context.

Convolutional Neural Networks (CNNs), particularly U-Net and its variants
such as ResNet-UNet \cite{Xiao}, UNet++ \cite {zhou}, and UNet3D \cite{huang}, have demonstrated strong performance in medical image segmentation by leveraging encoder-decoder architectures with skip connections to preserve spatial details \cite{unetsuccess}. To further enhance feature representation, attention mechanisms like squeeze-and-excitation \cite{sq_and_exit}, convolutional block attention \cite{cbam}, and dual attention modules were integrated into CNNs, enabling adaptive emphasis on informative regions and improving segmentation in low-contrast or structurally complex scenarios. These attention-enhanced CNNs have achieved notable success across various applications, including cardiac \cite{mri}, organ \cite{ct}, and lesion segmentation \cite{lesion}. However, their reliance on local convolutional operations limited the modeling of long-range dependencies, a critical factor in capturing global context in medical images \cite{chen}. The adoption of transformers in medical image segmentation introduced a paradigm shift by enabling global context modeling through self-attention. Unlike CNNs, which rely on local receptive fields, transformers capture long-range dependencies by computing pairwise interactions across all spatial tokens, crucial for identifying large or scattered anatomical structures. Vision Transformer (ViT) models and their variants, such as DINOv2 \cite{dino} and DeiT \cite{diet}, have demonstrated strong performance, even in data-limited settings. 

Medical-specific transformer architectures, such as Swin-Unet ~\cite{swinunet} and MedT \cite{medt}, leverage this capability to outperform CNNs across various modalities, including CT, MRI, and fundus imaging. However, representing images as 1D sequences often leads to low-resolution features and coarse segmentations that upsampling alone cannot resolve. Many studies have integrated attention mechanisms into CNN-based architectures to enhance long-range dependency modeling in medical image segmentation. Wang \textit{et al.} \cite{nonlocalblock} introduced a non-local block that computes responses at each spatial location as a weighted sum of features across the entire feature map, enabling global context modeling when inserted at multiple stages of the CNN backbone. Chen \textit{et al.} \cite{chen} proposed TransUNet, a hybrid architecture that utilizes CNNs for local feature extraction and Transformer blocks to capture global dependencies, followed by a U-Net decoder for segmentation. Schlemper \textit{et al.} \cite{attblock} designed attention gate modules for skip connections in U-Net-like architectures, allowing selective focus on salient features. Chang \textit{et al.} \cite{transclaw} presented TransClaw U-Net, which applies convolutional encoding followed by Transformer-based tokenization to model long-range context, with decoding handled by Claw U-Net's bottom-up structure. Xu \textit{et al.} \cite{levitunet} introduced LeViT-UNet, a lightweight model that combines multi-stage Transformer encoding with convolutional blocks and U-Net-style skip connections to balance global semantics and local spatial precision. Distinct from these approaches and drawing inspiration from the recent image classification study \cite{deepfake}.

We propose U-DFA, a hybrid unified DINOv2-UNet encoder-decoder architecture designed to integrate both local and global semantic features for medical image segmentation. It consists of three components: an encoder, a bottleneck, and a cascade decoder, with a focus on the encoder for extracting meaningful features. The encoder includes a head module with a Spatial Pattern Adapter (SPA) that runs parallel to the token embedding of a pre-trained, frozen DINOv2 Transformer. Each of the N intermediate stages contains one frozen Transformer block and a trainable Local-Global Fusion Adapter (LGFA), which fuses CNN and Transformer features using spatial and channel-wise attention. The decoder upsamples these features and incorporates multi-resolution skip connections from the SPA to refine spatial localization and boundaries. Our key contributions include: (1) A dual-path encoder that combines SPA with DINOv2 embeddings for effective local-global feature extraction.
(2) Design and integration of the LGFA module that enhances feature fusion at multiple levels. (3) An efficient configuration strategy that balances segmentation accuracy and model complexity, enabling practical deployment across diverse medical imaging tasks.

\vspace{-.5em}
\section{Method}
\subsection{Architecture Overview}
\begin{figure}[!t]
\includegraphics[width=1\textwidth]{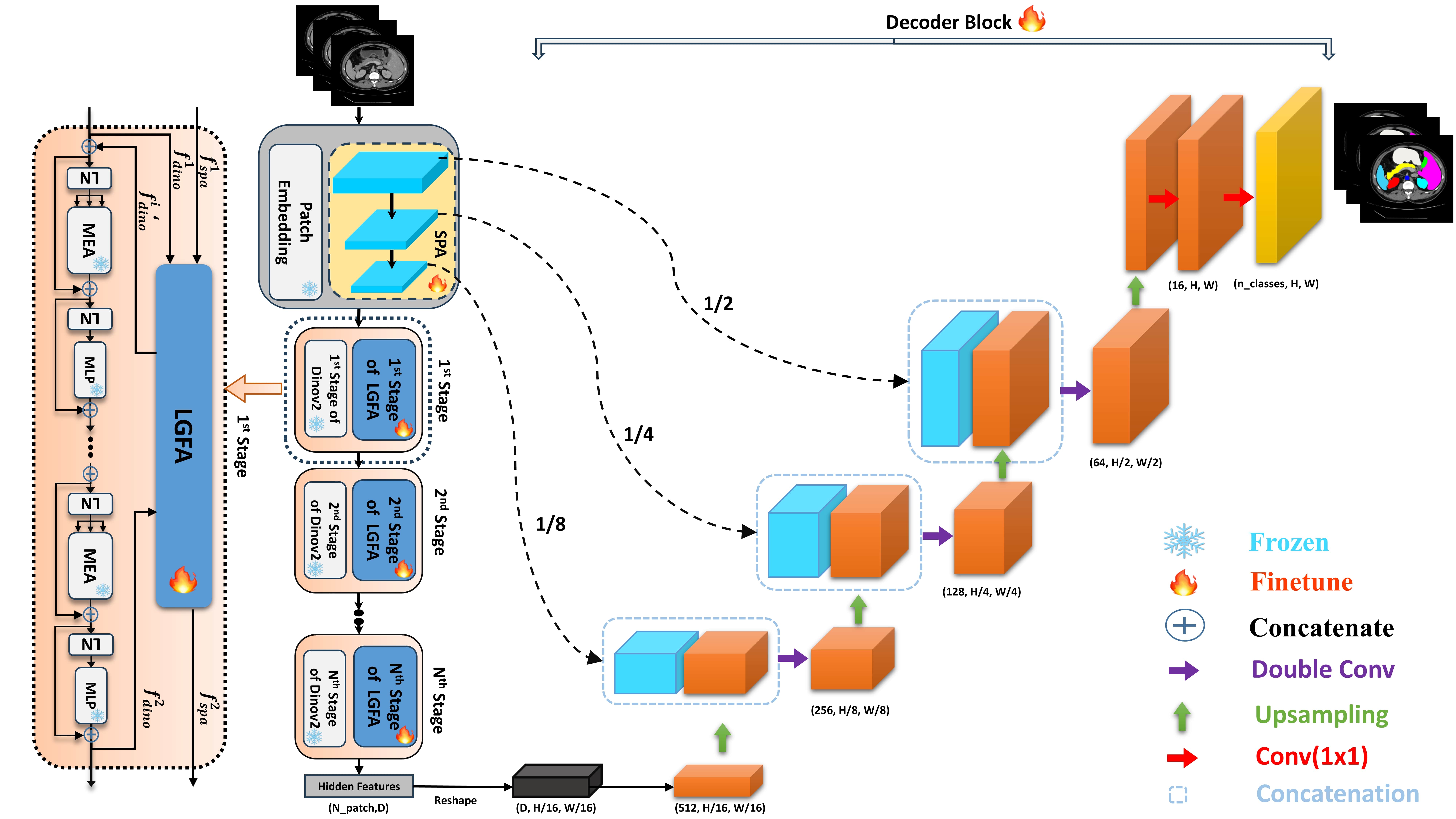}
\caption{Block diagram of the proposed U-DFA architecture.} \label{fig1}
\end{figure}

The proposed architecture of U-DFA is depicted in Fig. \ref{fig1}. Given an image \( I \in \mathbb{R}^{H \times W \times C} \) with spatial resolution \( H \times W \) and number of channels \( C=3\), it is fed into the encoder, which consists of a head and intermediate stages. The input image \(I\) is processed in parallel by the DINOv2 embedding layer and the SPA module in the head part of the proposed encoder. 

A DINOv2 embedding layer divides the image into \( P \times P \) non-overlapping patches and flattens them into sequential patches \(I_p \in \mathbb{R}^{K \times (P^2 \cdot C)}\), where \( K = H \cdot W / P^2 \) is the total number of patches. These flattened patches are projected into \textit{D}-dimensional embeddings and added with a positional embedding denoted as \(f^{1}_{dino} \in \mathbb{R}^{ \ (P^2 \cdot C)\times D}\) to retain the positional information. 

\begin{figure}[t]
\includegraphics[width=1\textwidth]{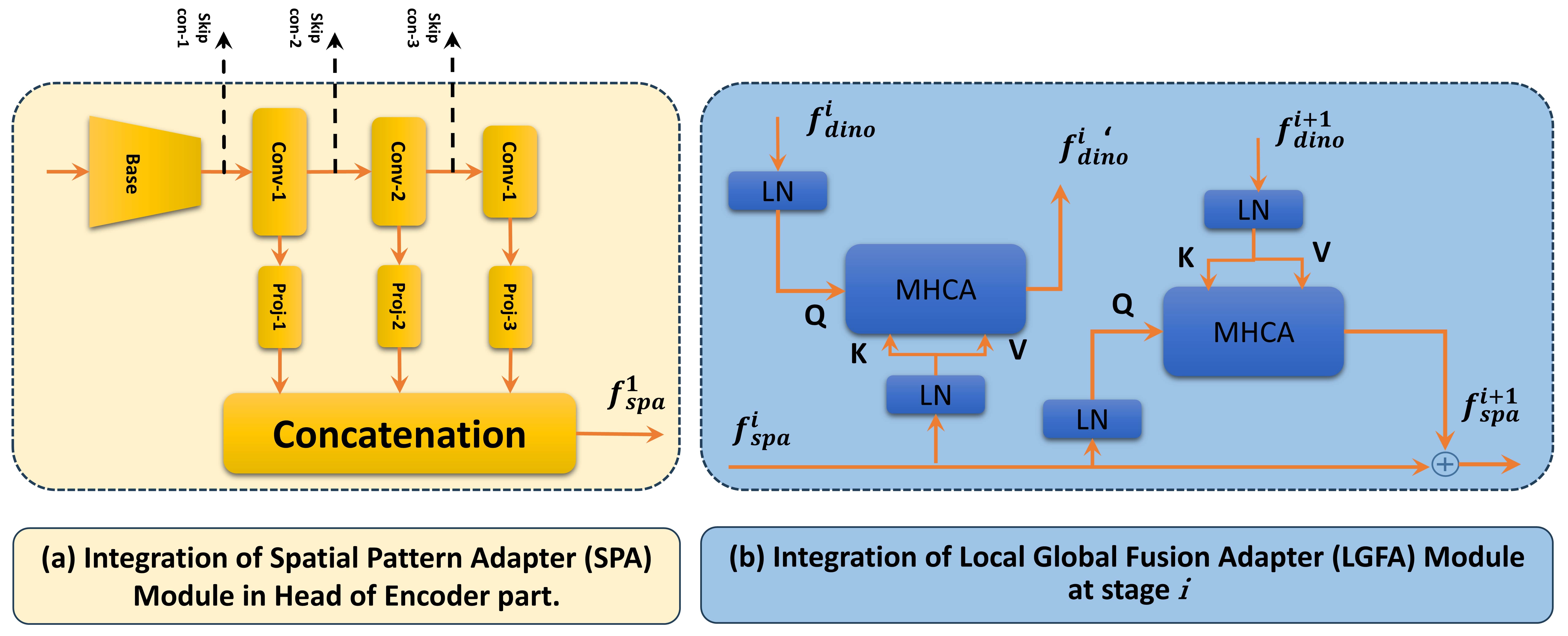}
\caption{Details of the SPA and LGFA module.} \label{fig2}
\end{figure}

A ResNet \cite{r50} inspired SPA module depicted in Fig. \ref{fig2} employs a standard CNN as a base network to extract the basic low-level feature maps using three similar Conv-BatchNorm-Relu blocks. Then the features are fed through three convolutional blocks to extract feature maps at different spatial resolutions, specifically at scales of \( 1/r_1 \), \( 1/r_2 \), and \( 1/r_3 \) relative to the input image size. These multi-scale feature maps are also utilized as skip connections in the decoder to facilitate the reconstruction of high-resolution outputs. Each feature map is then projected into a standard embedding dimension \(D\) using separate projection layers. The resulting vectors are concatenated to form a unified feature representation \( f^{1}_{spa} \in \mathbb{R}^{\left( \frac{HW}{r_1^2} + \frac{HW}{r_2^2} + \frac{HW}{r_3^2} \right) \times D} \). This representation enables the SPA module to aggregate rich, multi-scale local features, effectively capturing fine-grained spatial details.\newline

\noindent
\textbf{Transformers with N Stages} 
The extracted features \( f^{1}_{dino}\) and \( f^{1}_{spa}\) from the head part of the encoder are passed through the $1^{st}$ Stage of the encoder block. A pre-trained DINOv2-base backbone is utilized, comprising a total of \(L \) blocks, where each block consists of a Memory-Efficient Attention (MEA) and a Multi-Layer Perceptron (MLP) layer. \(N \) stages are formed by evenly grouping \(L \) blocks, with each stage containing \(L/N \) blocks of the DINOv2 and a single LGFA module for integration. We develop the LGFA interaction component for the SPA module, which facilitates the engagement of these features (e.g., \( f^{1}_{spa}\) in the $1^{st}$ Stage) with features from both the beginning and end of DINOv2 blocks at that Stage (e.g., \( f^{1}_{dino}\) and \(\ f^{2}_{dino}\) in the $1^{st}$ Stage).

Specifically, the interaction in the $i^{th}$ Stage begins with a Multi Head Cross Attention (MHCA) operation between \(\ f^{i}_{spa}\) and the features form the beginning of DINOv2 \(\ f^{i}_{dino}\), as shown in Fig. 2(b). During this process, the normalized DINOv2 features \( \widehat f^{i}_{dino}\) serves as the query while the normalized SPA features \(\widehat f^{i}_{spa}\) are used as both the key and value as follows,
\begin{equation}
    f^{i'}_{dino} = f^{i}_{dino} +
    \text{MHCA} \;(\widehat f^{i}_{dino},\; \widehat f^{i}_{spa},\; \widehat f^{i}_{spa})
\end{equation}
where \( f^{i'}_{dino} \) are the features from the first interaction of the LGFA module. These features are added element-wise with \(f^{i}_{dino}\) and then fed back into the DINOv2 blocks of the $i^{th}$ Stage resulting in \(f^{i+1}_{dino}\) features. This first interaction process injects the low-level features from the SPA module into the forward process of DINOv2 blocks. The second interaction in the $i^{th}$ Stage happens at the end of the DINOv2 blocks after getting \(f^{i+1}_{dino}\) features from the first interaction. The second interaction is performed between \( f^{i}_{\text{spa}} \) and \(  f^{i+1}_{dino} \) using the MHCA layer, where the role of key, query, and value is switched. We use normalized \( \widehat f^{i+1}_{dino} \) as the key and value, and normalized \(\widehat f^{i}_{spa} \) as the query as follows,
\begin{equation}
    f^{i+1}_{{spa}} = f^{i}_{\text{spa}} +
    \text{MHCA}\;(\widehat f^{i}_{spa},\; \widehat f^{i+1}_{dino},\; \widehat f^{i+1}_{dino})
\end{equation}
where \( f^{i+1}_{spa} \) represents the updated low-level features that will interact with the new features from the DINOv2 blocks \( f^{i+1}_{dino} \) in the subsequent stage. Consequently, the encoded features will be further enhanced during the dual fusion process at the end of each stage. 

After the extraction of \( f^{N+1}_{spa} \) features through \( N \) stages of the encoder block, the features are forwarded to the bottleneck, where the spatial dimensions are reshaped from \( HW / P^2 \) to \( (H / P) \times (W / P) \) resolution. A single \(1 \times 1\) convolution is applied to reduce the channel dimension of the reshaped features to match the number of target classes, and then the output is forwarded to the decoder block. Finally, the feature map is upsampled to the original spatial resolution \(H \times W\) using bilinear interpolation, followed by a three-stage \texttt{DoubleConv} block. Each stage of the \texttt{DoubleConv} block comprises two consecutive \(3\times3\) convolutional layers, each followed by Batch Normalization and ReLU activation functions. To enhance feature maps and preserve spatial details, skip connections from the SPA module are incorporated at each stage to prevent spatial loss. Subsequently, two \(1 \times 1\) convolutional layers are applied at the end to predict the final segmentation mask.
\vspace{-.5em}
\section{Experiments}
\subsection{Benchmark Datasets}
\textbf{Synapse multi-organ segmentation dataset:}
The Synapse multi-organ segmentation dataset, released as part of the abdominal organ segmentation challenge "Beyond the Cranial Vault (BTCV)", serves as a standardized benchmark for tasks involving medical image segmentation. It consists of 30 abdominal computed tomography (CT) volumes, encompassing a total of 3,779 axial contrast-enhanced clinical CT slices with an original 512 \( \times \) 512 resolution. Each volume includes manual annotations for eight abdominal organs. Following previous works \cite{chen,attunet,swinunet}, we adopt the same dataset splitting strategy, using 18 volumes for training and 12 volumes for testing. We downsample all images to a resolution of 224 \( \times \) 224. For performance evaluation, we employ the average Dice Similarity Coefficient (DSC) and average Hausdorff Distance (HD) as evaluation metrics. \newline
\textbf{Automated Cardiac Diagnosis Challenge:} The ACDC is a dataset of 100 patients used for 3D volumetric MRI scans. Each patient's MRI image includes labeled regions for the right ventricle (RV), left ventricle (LV), and myocardium (Myo). A dataset splitting strategy in line with \cite{chen,swinunet} is followed, and the segmentation accuracy is evaluated using the Dice metric and the average Intersection over Union (IoU). The dataset is divided into 70\% training samples, 10\% validation samples, and 20\% testing samples.
\vspace{-.5em}
\begin{table*}[b]
\centering
\caption{Comparison of different methods on Synapse dataset using (average dice score, average Hausdorff Distance (HD), and Dice score (\%) in each class)}
\label{tab:synapse_results}
\renewcommand{\arraystretch}{1.2}
\resizebox{\textwidth}{!}{
\begin{tabular}{l|c c|cccccccccc}
\hline
\textbf{Methods} & \textbf{DSC}$\uparrow$ & \textbf{HD}$\downarrow$ & \textbf{Aorta} & \textbf{Gallbladder} & \textbf{Kidney(L)} & \textbf{Kidney(R)} & \textbf{Liver} & \textbf{Pancreas} & \textbf{Spleen} & \textbf{Stomach} \\
\hline
R50 ViT~\cite{chen}    & 71.29 & 32.87 & 73.73 & 55.13 & 76.29 & 72.20 & 91.51 & 45.99 & 81.99 & 73.95 \\
R50 U-Net~\cite{chen}  & 74.68 & 36.87 & 87.74 & 63.66 & 80.60 & 78.19 & 93.74 & 56.90 & 85.87 & 74.16 \\
R50 Att-UNet~\cite{chen} & 75.57 & 36.97 & 55.92 & 63.91 & 79.20 & 72.71 & 93.56 & 49.37 & 87.19 & 74.95 \\
U-Net~\cite{unettab}       & 76.85 & 39.70 & 89.07 & 69.72 & 77.77 & 68.60 & 93.43 & 53.98 & 86.67 & 75.58 \\
Att-UNet~\cite{attunet}  & 77.77 & 36.02 & 89.55 & 68.88 & 77.98 & 71.11 & 93.57 & 58.04 & 87.30 & 75.75 \\
TransUNet~\cite{chen} & 77.48 & 31.69 & 87.23 & 63.13 & 81.87 & 77.02 & 94.08 & 55.86 & 85.08 & 75.62 \\
TransClaw U-Net~\cite{transclaw} & 78.09 & -- & 85.87 & 61.38 & 84.83 & 79.36 & 94.28 & 57.65 & 87.74 & 73.55 \\
LeVit-UNet-384~\cite{levitunet} & 78.53 & 16.84 & 87.33 & 62.23 & 84.61 & 80.25 & 93.11 & 58.07 & 88.86 & 72.76 \\
Swin-Unet~\cite{swinunet} & 79.13 & 21.55 & 85.47 & 66.53 & 83.28 & 79.61 & 94.29 & 56.58 & 90.66 & 76.60 \\
MISSFormer~\cite{missformer} & 81.96 & -- & 86.99 & 68.65 & 85.21 & 82.00 & 94.41 & \textbf{65.67} & 91.92 & 80.81 \\
DSGA-Net \cite{dsganet}  & 81.24 & 20.91 & 88.21 & \textbf{70.87} & 82.67 & 82.31 & 95.76 & 58.49 & 90.87 & 80.74\\
RotU-Net \cite{routnet} & 82.15 &26.95 &89.03 &70.51 & 82.74 & 81.79 & 95.29 & 64.92 & \textbf{91.92} & 80.81 \\
\textbf{Ours}  & \textbf{82.25} & \textbf{15.27} & \textbf{89.85} & 69.02 & \textbf{85.58} & \textbf{83.11} & \textbf{95.92} & 61.17 & 89.99 & \textbf{83.35} \\
\hline
\end{tabular}}
\end{table*}
\begin{table}[t]
\centering
\caption{Performance comparison with different methods on the ACDC dataset. }
\label{tab:acdc_results}
\renewcommand{\arraystretch}{1}
\begin{tabular}{l|c|ccc}
\hline
\textbf{Methods} & \textbf{DSC} & \textbf{RV} & \textbf{Myo} & \textbf{LV} \\
\hline
R50 U-Net \cite{chen}     & 87.55 & 87.10 & 80.63 & 94.92 \\
R50 Att-UNet \cite{chen}   & 86.75 & 87.58 & 79.20 & 93.47 \\
R50 ViT    \cite{chen}    & 87.57 & 86.07 & 81.88 & 94.75 \\
UNETR \cite{unetr}         & 88.61 & 85.29 & 86.52 & 94.02 \\
TransUnet \cite{chen}      & 89.71 & 88.86 & 84.53 & 95.73 \\
DAE-Former[44] & 89.78 & 89.91 & 84.38 & 95.04 \\
Swin-UNet \cite{swinunet}  & 90.00  & 88.55 & 85.62 & 95.83 \\
\textbf{Ours} & \textbf{90.46} & 87.85 &  \textbf{87.53} & \textbf{96.01} \\
\hline
\end{tabular}
\end{table}

\subsection{Experimental Results:}
The implementation is carried out in Python 3.10 using the PyTorch 2.6.0 framework. The hardware setup consists of an NVIDIA RTX 3090 GPU with 24 GB of VRAM. We resized the input images to 224 \( \times \) 224, with a batch size of 12, during the training process. To enhance the robustness of the model, data augmentation techniques such as random flipping, rotation, and intensity randomization were applied. Furthermore, a pretrained DINOv2-base backbone is employed, which was kept frozen throughout the training. We use the Adam optimizer with a weight decay of \(1\times 10^{-4}\). Finally, the total loss function is the sum of Dice loss and cross-entropy loss with equal weightage.\newline
\textbf{Synapse Dataset:}
A comparative analysis of the proposed method against several state-of-the-art (SOTA) segmentation frameworks on the Synapse dataset is summarized in Table \ref{tab:synapse_results}. Our model achieves the highest average DSC of 82.25\%, outperforming all existing methods, including RotU-Net (82.15\%), MISSFormer (81.96\%), and DSGA-Net (81.24\%), and achieves a 15.27\% HD score compared to the previous SOTA methods. 
In organ-wise evaluation, our method achieves the best performance on five out of eight organs, including Aorta (89.85\%), Kidney(L) (85.58\%), Kidney(R) (83.11\%), Liver (95.92\%), and Stomach (83.35\%). This demonstrates the robustness of our model across anatomically diverse and complex organ structures. While DSGA-Net attains the highest Dice score for the Gallbladder (70.87\%) and MISSFormer and RotUnet perform best on the Pancreas (65.67\%) and Spleen (91.92\%), respectively, our method remains highly competitive across these challenging organs. Furthermore, our approach achieves this performance using only 33\% of the model parameters for fine-tuning instead of training the entire model end-to-end, demonstrating both its efficiency and effectiveness. \newline
\textbf{ACDC Dataset:} Table \ref{tab:acdc_results} contrasts and compares the results on the ACDC dataset. The proposed method out performs the SOTA, achieving the highest overall average DSC score of 90.46\% along with superior segmentation accuracy for Myo (87.53\%) and LV (96.01\%). Pure transformer approaches, including UNETR, DAE-Form, and Swin-Unet, achieve an average DSC score of 88.61\%, 89.78\%, and 90.00\%, respectively. The improvements highlight our model’s strong generalization and feature representation capabilities for cardiac MRI segmentation tasks.

\subsection{Ablation Study:}
To thoroughly evaluate the proposed U-DFA method under different settings, ablation studies were performed, including input resolution and number of LGFA modules, as discussed below: \newline
\textbf{Effect of image size and number of LGFA modules:}
Two input image resolutions, \(224 \times 224\) and \(308 \times 308\), are used for the Synapse dataset to evaluate our method and examine the effects of changing the image size. The results are presented in Table \ref{tab:ablation_combined}, which shows that increasing the input image size leads to a slight improvement in HD. However, changing the image size from  \(224 \times 224\) to \(308 \times 308\) results in an increase in computational cost, as the patch size remains the same (i.e., \(14 \times 14\)). To investigate the effect of the number of LGFA modules in the encoder, we conducted an ablation study on the Synapse dataset by varying the number of modules: 2, 3, and 6, using an input resolution of \(224 \times 224\).
\begin{figure}[htbp]
\centering
\includegraphics[width=0.9\textwidth, height=5.5cm]{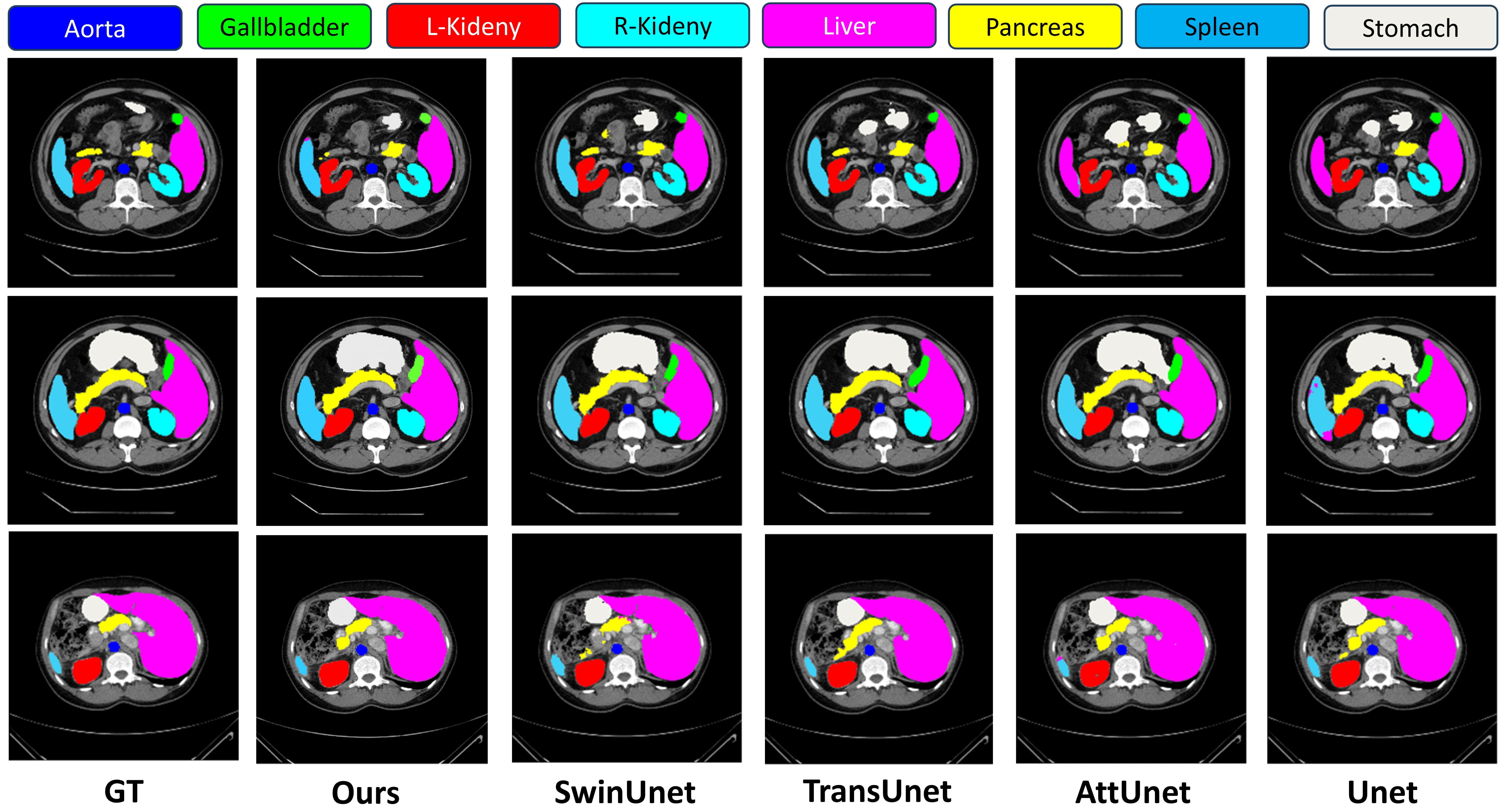}
\caption{Results on Synapse multi-organ CT
dataset and comparison of our method with others.
} \label{fig3}
\end{figure}
\begin{table}[htbp]
\centering
\caption{Ablation study on input size and number of LGFA modules on Synapse dataset.}
\label{tab:ablation_combined}
\renewcommand{\arraystretch}{1.2}
\resizebox{\textwidth}{!}{
\begin{tabular}{c|c|cc|cccccccccc}
\hline
\textbf{Input Size} & \textbf{No. of LGFA} & \textbf{DSC}$\uparrow$ & \textbf{HD}$\downarrow$ & \textbf{Aorta} & \textbf{Gallbladder} & \textbf{Kidney(L)} & \textbf{Kidney(R)} & \textbf{Liver} & \textbf{Pancreas} & \textbf{Spleen} & \textbf{Stomach} \\
\hline
224 & 2 & 82.09 & 18.97 & 89.16 & 68.61 & 85.49 & 80.41 & 95.73 & 64.76 & 90.08 & 82.49 \\
224 & 3 & 82.25 & 15.27 & 89.85 & 69.02 & 85.58 & 83.11 & 95.92 & 61.17 & 89.99 & 83.35 \\
224 & 6 & 82.67 & 19.76 & 89.90 & 70.10 & 84.86 & 81.51 & 95.63 & 66.94 & 90.20 & 82.21 \\
308 & 3 & 82.37 & 15.42   & 90.25 & 69.37 & 83.26 & 81.76 & 96.04 & 65.05 & 90.16 & 83.04 \\
\hline
\end{tabular}
}
\end{table}

While the overall DSC remained relatively similar, the HD showed notable variation. The configuration with 3 LGFA modules achieved the lowest HD of 15.27\%, indicating better boundary delineation. In comparison, 2 and 6 modules resulted in HD scores of 18.97\% and 19.76\%, respectively, suggesting under-utilization and potential overfitting. These results highlight that using 3 LGFA modules offers the best trade-off between segmentation accuracy and boundary precision, providing an optimal balance of model complexity and performance. We also evaluated our method on the \textbf{LUNA16} dataset for lung segmentation, achieving an average DSC of 96.54\% and IoU of 90.79\%. These results demonstrate the robustness of our approach in accurately segmenting lung regions, further validating its generalizability.

\section{Conclusions}
In this paper, we present a robust and scalable method, U-DFA, for medical image segmentation leveraging the Unet and DINOv2 architectures. With the help of these two distinct architectures, we have effectively fused local and global features. To further improve cross-scale feature interaction, we introduced the LGFA module, which enhances feature fusion across different levels of the network. We have evaluated our proposed method across multiple datasets and compared the results with the current state of the art in the domain. Our method achieves state-of-the-art performance on the Synapse and ACDC datasets with only 33\% trainable model parameters. The results demonstrate the superiority of our proposed method and reflect its suitability for deployment in practical scenarios. As a next step, we aim to adapt U-DFA for 3D volumetric segmentation tasks and explore the use of prompt-driven and zero-shot learning approaches to enhance further the framework's flexibility across unseen medical imaging scenarios.

\end{document}